# Observation of Weyl nodes and Fermi arcs in TaP

N. Xu[1,2,*,§], H. M. Weng[3,4,*], B. Q. Lv[1,3], C. E. Matt[1], J. Park[1], F. Bisti[1], V. N. Strocov[1], D. Gawryluk[5], E. Pomjakushina[5], K. Conder[5], N. C. Plumb[1], M. Radovic[1], G. Autès[6,7], O. V. Yazyev[6,7], Z. Fang[3,4], X. Dai[3,4], T. Qian[3], J. Mesot[1,2,8], H. Ding[3,4,§] and M. Shi[1,§]

[1] *Swiss Light Source, Paul Scherrer Institut, CH-5232 Villigen PSI, Switzerland*

[2] *Institute of Condensed Matter Physics, École Polytechnique Fédérale de Lausanne, CH-1015 Lausanne, Switzerland*

[3] *Beijing National Laboratory for Condensed Matter Physics and Institute of Physics, Chinese Academy of Sciences, Beijing 100190, China*

[4] *Collaborative Innovation Center of Quantum Matter, Beijing, China*

[5] *Laboratory for Developments and Methods, Paul Scherrer Institut, CH-5232 Villigen, Switzerland*

[6] *Institute of Theoretical Physics, École Polytechnique Fédérale de Lausanne, CH-1015 Lausanne, Switzerland*

[7] *National Center for Computational Design and Discovery of Novel Materials MARVEL, Ecole Polytechnique Fédérale de Lausanne (EPFL), CH-1015 Lausanne, Switzerland*

[8] *Laboratory for Solid State Physics, ETH Zürich, CH-8093 Zürich, Switzerland*

\* These authors contributed equally to this work.

§ E-mail: nan.xu@psi.ch , dingh@iphy.ac.cn , ming.shi@psi.ch

**Abstract**

A Weyl semimetal possesses spin-polarized band-crossings, called Weyl nodes, connected by topological surface arcs. The low-energy excitations near the crossing points behave the same as massless Weyl fermions, leading to exotic properties like chiral anomaly. To have the transport properties dominated by Weyl fermions, Weyl nodes need to locate nearly at the chemical potential and enclosed by pairs of individual Fermi surfaces with nonzero Fermi Chern numbers. Combining angle-resolved photoemission spectroscopy and first-principles calculation, here we show that TaP is a Weyl semimetal with only single type of Weyl fermions, topologically distinguished from TaAs where two types of Weyl fermions contribute to the low-energy physical properties. The simple Weyl fermions in TaP are not only of fundamental interests but also of great potential for future applications. Fermi arcs on the Ta-terminated surface are observed, which appear in a different pattern from that on the As-termination in TaAs and NbAs.

# Introduction

A Weyl semimetal (WSM) has recently attracted great attentions as a novel quantum state in which spin non-degenerate band crossings near the chemical potential are topologically unavoidable [1, 2]. The band crossing points, called Weyl nodes, always appear in pair(s) with opposite chirality and the bands near the Weyl nodes disperse linearly in three-dimensional (3D) momentum space that can be described by reduced two-component massless Dirac equations [3]. An isolated Weyl node is a sink or a source of Berry curvature whose chirality corresponds to a topological charge [4, 5] and exhibits chiral anomaly [6-8] that can give rise to various exotic transport phenomena such as negative magnetoresistance (MR) [7], non-local quantum oscillation [9], and quantum anomalous Hall effect [5]. In addition WSM hosts topologically non-trivial surface states forming arcs, so-called Fermi arcs, extending the classification of topological phases beyond insulators. The WSM state needs no additional symmetry protection other than translational symmetry in the crystal and therefore can be more robust than the other topological semimetal states in which additional crystal symmetry are needed [10-17].

Various materials have been proposed as candidates for the WSM based on time-reversal symmetry breaking [1, 5, 18] but the experimental realization remains doubtful [19-28]. Recently the WSM based on inversion symmetry breaking has been predicted to be realized in nonmagnetic and non-centrosymmetric transition metal monoarsenides/monophosphides [29, 30] and indeed the key features of the WSM including the Fermi arcs [31-34] and the Weyl nodes [35] and the negative MR [36-38] have been observed in TaAs. These monoarsenides/monophosphides family has two types of the Weyl nodes: the W1 Weyl nodes in a pair well separated in momentum space off the $k_z = 0$ plane and the W2 Weyl nodes very close to each other in $k_z = 0$ plane. In TaAs, W1 and W2 are not related with each other by any symmetry, which makes W1 and W2 have different velocity matrix and locate at different energy level. However, the Fermi Chern number ($C_{FS}$) [10] (net topological charge enclosed by a Fermi surface) of a Fermi surface determines the low energy physical properties of WSM. In TaAs, each of the W1 and W2 Weyl nodes is enclosed by a single Fermi surface under experimental carrier density [31, 34-36]. This makes the transport properties quite complex with mixed contributions from both types of Weyl nodes.

The realization of WSMs with well-separated Weyl nodes of single type sitting at a chemical potential is crucial for clearly revealing the fundamental physics and practical applications of the novel WSM states since it is an ideal platform having the simplest Weyl nodes in WSMs [21]. A promising candidate is TaP in which large and unsaturated negative MR has been observed [39-41]. Here, we present the electronic structure of TaP demonstrating the complete feature of the WSM states. In the bulk, pairs of the W1 Weyl nodes well-separated in momentum space have been found to locate at the chemical potential while the poorly-separated ones W2 are 60 meV below the chemical potential, with pairs of opposite chiral W2 enclosed by one Fermi surface. This leads to only a single type Weyl fermions dominating the exotic low energy physical properties, such as observed magneto-transport phenomena and quantum oscillation. The Fermi arcs on the Ta-terminated surface have also been observed, with an entirely different pattern from that on the As-terminated surface observed in TaAs and NbAs [31-34]. Our results illustrate that the Fermi arc states strongly depend on the details of surface conditions, which has to be considered in the transport and spectroscopic properties of a system where surface contribution is non-neglectable, such as thin film and non-local quantum oscillations.

## Results
### Weyl semimetal state in TaP single crystal

TaP has a body-centred-tetragonal structure without inversion symmetry (space group I4$_1$md), depicted in Fig. 1a. The lattice parameters obtained from our x-ray diffraction measurement at room temperature are $a = b = 3.32$ Å and $c = 11.34$ Å (see Supplementary Figure 1), consistent with a previous report [42]. An *Ab initio* calculation predicts twelve pairs of the Weyl nodes in the Brillouin zone (BZ) (Fig. 1b) where eight pairs locate at $k_z \sim \pm 0.6 \, \pi/c'$ (named W1) where $c' = c/2$ and the other four reside in the $k_z = 0$ plane (W2) [16]. Figure 1c illustrates the band structure near a pair of Weyl nodes where the color of spheres represents the chirality. The energy bands near Weyl nodes disperse linearly in all the directions in momentum space and satisfy following conditions. (1) Along the cut passing through a pair of Weyl nodes (cut$_x$ in Fig. 1c), two spin non-degenerated bands cross each other twice and disperse linearly near the crossing points (Fig. 1d). (2) The band along a cut perpendicular to cut$_x$ (cut$_y$ or cut$_z$ in Fig. 1c) passes through one Weyl node and exhibits a single Dirac cone dispersion, seen in Fig. 1d. (3) The band in the $k_x$-$k_y$ plane (cut$_{x-y}$ in Fig. 1c) containing a pair of the Weyl nodes has a double-Dirac-cone dispersion, as depicted in Fig. 1e.

**Bulk electronic structure of TaP**

The bulk electronic structure of TaP has been characterized by angle-resolved photoemission spectroscopy (ARPES) with photon energy in the soft x-ray region, which has high bulk sensitivity and $k_z$ resolution [43-44]. Figure 2a presents the ARPES spectra in the $k_y$-$k_z$ plane (the shadow plane in Fig. 1b), acquired with $h\nu$ = 300-700 eV. Both the Fermi surface (FS) and the band disperses along $k_z$ show a periodicity of $2\pi/c'$, indicating a bulk origin. The measured bands are in good agreement with the bulk band structure calculated with spin-orbit coupling (SOC), as shown in Fig. 2b. Due to the absence of inversion symmetry and the strong SOC in TaP, the spin degeneracy is lifted and the bulk states split into two bands, which are clearly observed in the ARPES results and band calculations (Fig. 2b).

**Observation of Weyl nodes in TaP**

With the above conditions, we have identified the Weyl nodes W1 and characterized them. Figure 2d shows that the FS splits into two small pockets sitting aside the $k_{x/y}$ = 0 plane, which is fully consistent with the band structure of Weyl nodes (Fig. 1e). The ARPES intensities and the corresponding curvatures along $cut_x$ shown in Fig. 2e,f and Fig. 2g,h (for $k_z$ = ±0.58 $\pi/c'$ as seen from Fig. 2i) display that two bands cross each other twice with the Weyl nodes at $k_x$ = ±0.03 $\pi/a$, in good agreement with our calculation. The band bottom of the electron-like band is slightly below $E_F$, which is enhanced with photons of 478 eV (Fig. 2g,h) due to the matrix element effect. The band structures along $cut_y$/$cut_z$ exhibit the Weyl nodes locating at $k_y$ = 0.54 $\pi/a$ and $k_z$ = 0.58 $\pi/c'$, also remarkably consistent with the calculation as seen in Fig. 2j-m. The band evolutions along $k_y$/$k_z$ further verify the 3D nature of this double Dirac cone dispersion (see Supplementary Figure 2). Within the experimental error the Weyl nodes W1 locate at the chemical potential and the distance between opposite chirality nodes in a pair is 0.06 $\pi/a$.

The other sort of nodes W2 locating at the $k_z$ = 0 plane has also been identified. Figure 3b shows the FS map recorded at $h\nu$ = 455 eV corresponding to $k_z$ = 0. Because the chemical potential is 60 meV above the Weyl nodes W2, the FS shows a single pocket at the middle points of the BZ boundary, in contrast to the FS of W1 that forms two small pockets slightly off the mirror plane. W2 residing below the chemical potential allow to us observe the whole band dispersion near the Weyl nodes, including the upper branches of Dirac cones. A double Dirac cone dispersion displaying in the band structures along the $cut_x$ and a single Dirac cone dispersion with a small Fermi velocity along the $cut_y$ and $cut_z$ are shown in Fig. 3c-i. The evolutions of the experimentally determined band structure along $k_y$ and $k_z$ directions (see Supplementary Figure 3) further confirmed the Dirac cone dispersions in all three directions of momentum space at the Weyl nodes W2. The full consistency between the experiments and the first-principles calculations unambiguously establishes the Weyl nodes W2 at ($\pm 0.01$ $\pi/a$, $\pm 1.03$ $\pi/a$, 0) with a separation of 0.02 $\pi/a$ in a pair. These clear show that every pair of W2 with opposite chirality or topological charge is enclosed by single Fermi surface with a zero $C_{FS}$ [10, 45], and such Fermi surface is topologically trivial and its low energy physics is not associated with Weyl fermions [46].

**Fermi arcs on Ta-terminated surface in TaP**

Fermi arcs appearing at the surface that connect the surface projections of the Weyl nodes in pairs are another key feature of the WSM, and induce exotic transport properties like non-local quantum oscillation. The surface sensitivity of ARPES with vacuum ultra-violet light (VUV-ARPES) enables us to distinguish the surface states from the bulk states. In Fig. 4a, comparison between angle-integrated core level spectra taken with soft x-ray and VUV light reveals that Ta- and P-terminated surface states coexist; Two additional P-*2p* peaks and the broadening and position change of the Ta-*4f* peak in the VUV spectrum indicate the P-terminated surface states and Ta-terminated ones, respectively.

The FS and band structure measured at $h\nu$ = 50 eV well agree with the calculation of the surface states on the Ta-terminated (001) surface (Fig. 4b-f). Our calculation reveals that on the Ta-terminated surface along the $\bar{\Gamma}$-$\bar{X}$ direction the Fermi arcs between the pairs of W1 nodes are very short, thus it is difficult to resolve them. On the other hand, along the $\bar{\Gamma}$-$\bar{Y}$ direction the Fermi arcs connect the W1 nodes in adjacent BZs, marked as $Ta_{3-4}$ (Fig. 4c). Consistent with the calculation (Fig. 4c), the FS map determined from ARPES measurements (Fig. 4b) clearly shows the Fermi arc

Ta$_{3-4}$ terminating at the projection of the W1 nodes at (0.03 π/a, 0.54 π/a) and (0.03 π/a, 1.46 π/a) on the surface. At a higher binding energy ($E_B$ = -0.1 eV), the end point of the Fermi arc Ta$_{3-4}$ goes away from the Weyl nodes (Fig. 4d), consistent with the dispersion of the bulk Weyl cones. The intensity of the surface states associated to the P-terminated surface is weak and only two hole-like pockets centered at the $\bar{\Gamma}$ point (P$_{1-2}$ in Fig. 4e) are visible in the FS map (Fig. 4b) and in the band structure (Fig. 4f, also see Supplementary Figure 4). The data measured with VUV lights have been confirmed to originate because the obtained FS does not change when varying the incident photon energy (Fig. 4g). The topology of the Ta-terminated Fermi arc states observed in TaP is totally different from the formerly reported As-terminated ones in TaAs and NbAs [31-34]. Our results illustrate that the Fermi arc states strongly depend on the details of surface conditions, which has to be considered for interpolating the transport and spectroscopic properties when surface contribution is non-neglectable.

## Discussion

In Fig. 5, the schematic diagrams shows a comparison of the band dispersions near the Weyl nodes in TaAs and TaP, based on the ARPES results and first-principles calculations on TaAs from Ref. [35] and on TaP in this work (see Supplementary Figure 5). Different from TaAs in which each of the W1 and W2 Weyl nodes is enclosed by a single Fermi surface ($C_{FS}$ = ±1) (Fig. 5a), the W2 Weyl nodes sink far below $E_F$ in TaP, with pairs of opposite chiral nodes enclosed by one Fermi surface (Fig. 5b). Such a Fermi surface has a topologically trivial Fermi Chern number ($C_{FS}$ = 0), makes TaP qualitatively different from TaAs in the topological aspect.

All the exotic transport properties of WSM, including negative MR and non-local transport/quantum oscillation, are due to the Weyl nodes with opposite chirality well separated in the momentum space and each node being enclosed by a single Fermi surface. Considering that W1 nodes in TaP is well separated and nearly sitting on the Chemical potential and pairs of W2 nodes are enclosed by one Fermi surface, TaP becomes an ideal WSM [21] with only one type of Weyl nodes contributes to its anomalous low energy physics. This makes it a very unique platform to study the underlying mechanism of the non-saturated negative magneto resistance [39-41] and quantum phase transition [47] observed only in TaP in the whole transition-metal monophosphides family.

In the transition metal monoarsenides Weyl semimetals, the Fermi arcs have been observed only at the one side of the surfaces, As-terminated, and the experimental observation of the arcs on the other side needed for forming a closed Fermi surface loop is missing. We have observed the Fermi arcs on the Ta-termination surface, with a different pattern from that on the As-termination in TaAs and NbAs. Our results illustrate that the surface termination effect on the Fermi arcs has to be taken into account for understanding the properties of WSMs when surface contribution is non-neglectable.

## Methods

**Sample synthesis**

Single crystals of TaP were grown by a chemical vapor transport in a temperature gradient 850 °C→950 °C, using 0.6 g of polycrystalline TaP as a source and iodine as a transport agent with a concentration of 12.2 mg/cm$^3$. Polycrystalline TaP was synthesized by a solid state reaction using elemental niobium and red phosphorus of a minimum purity 99.99 %. The respective amounts of starting reagents were mixed and pressed into pellets in the He-glove box, and annealed in the evacuated quartz ampule at 1050 °C for 60 hours. The laboratory x-ray diffraction measurements, which were done at room temperature using Cu Kα radiation on Brucker D8 diffractometer, have proven that the obtained crystals are single phase with the tetragonal structure of space group I4$_1$md.

**Angle-resolved photoemission spectroscopy**

Clean surfaces for ARPES measurements were obtained by cleaving TaP samples *in situ* in a vacuum better than 5 × 10$^{-11}$ Torr. VUV-ARPES measurements were performed at the Surface and Interface (SIS) beamline at the Swiss Light Source (SLS) with a Scienta R4000 analyser. Soft x-ray ARPES measurements were performed at the Advanced Resonant Spectroscopies (ADRESS) beamline at SLS with a SPECS analyser [42], and data were collected at $T$ = 10 K using circular-polarized light with an overall energy resolution on the order of 50-80 meV for FS mapping and 40-60 meV for high-resolution cuts measurements.

**Calculation Methods**

First-principles calculations were performed using the OpenMX [48] software package. The pseudo atomic orbital basis set with Ta9.0-s2p2d2f1 and P9.0-s2p2d1 was taken. The pseudo-potentials for Ta and P were generated with the exchange-correlation functional within generalized gradient approximation parameterized by Perdew, Burke and Ernzerhof [49]. The sampling of BZ (10 × 10 × 10 $k$-grid) was used. The setting of these parameters was tested to describe the electronic structure accurately. The lattice constants $a = b$ = 3.318 Å, $c$ = 11.363 Å and atomic sites have been used in the calculations. For (001) surface state calculation, a slab spanned by lattice $a$ and $b$ with thickness of seven unit cells along $c$ lattice is used. The slab is separated by vacuum layer of 12 Å to avoid interaction between slabs in neighbouring unit cell. Such self-consistent slab calculation can largely consider the surface relaxation and modification to potential in real sample [31]. The eigen wave function is projected onto the outmost eight layers of atoms (corresponding to one unit cell along $c$ lattice) so that its spatial distribution can be well characterized [31].

# End Notes

**Acknowledgements**

We acknowledge the help in plotting figures from Weilu Zhang. This work was supported by NCCR-MARVEL funded by the Swiss National Science Foundation, the Sino-Swiss Science and Technology Cooperation (No. IZLCZ2138954), and the Swiss National Science Foundation (No. 200021-137783), the Ministry of Science and Technology of China (No. 2013CB921700, No. 2015CB921300, No. 2011CBA00108, and No. 2011CBA001000), the National Natural Science Foundation of China (No. 11474340, No. 11422428, No. 11274362, and No. 11234014), the Chinese Academy of Sciences (No. XDB07000000). G.A. and O.V.Y. acknowledges support by the Swiss NSF (grant No. PP00P2_133552), ERC project "TopoMat" (grant No. 306504).

**Author contributions**

N.X., H.D. and M.S. conceived the experiments. N.X. and B.Q.L. carried out the experiments with assistance from C.E.M., J.P., F.B., V.N.S, N.C.P. and M.R.; H.M.W., Z.F. and X.D. performed *ab* initio calculations. N.X., Q.T., H.D and M.S. analysed the data with valuable feedback from G.A., O.V.Y. and J.M.; E.P., D.G. and K.C. synthesized and characterized the single crystals; N.X., H.M.W., H.D. and M.S. wrote the manuscript. All authors discussed the results and commented on the manuscript.

**Competing financial interests**

The authors declare no competing financial interests.

# Figure Legends

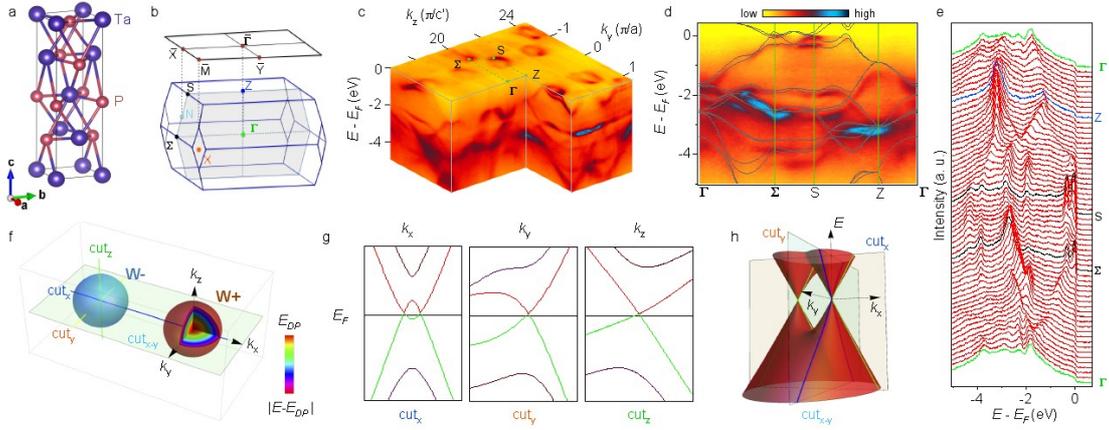

**Figure 1. Pairs of Weyl nodes in TaP. a**, Crystal structure of TaP. **b**, Bulk and surface BZ of TaP with high-symmetry points labelled. **c**, Illustration of a pair of Weyl nodes in 3D momentum space. **d**, Band structure passing through Weyl node along $cut_x$, $cut_y$ and $cut_z$. **e**, Schematic band dispersions of a pair of Weyl nodes in the $k_x$-$k_y$ plane of the momentum space.

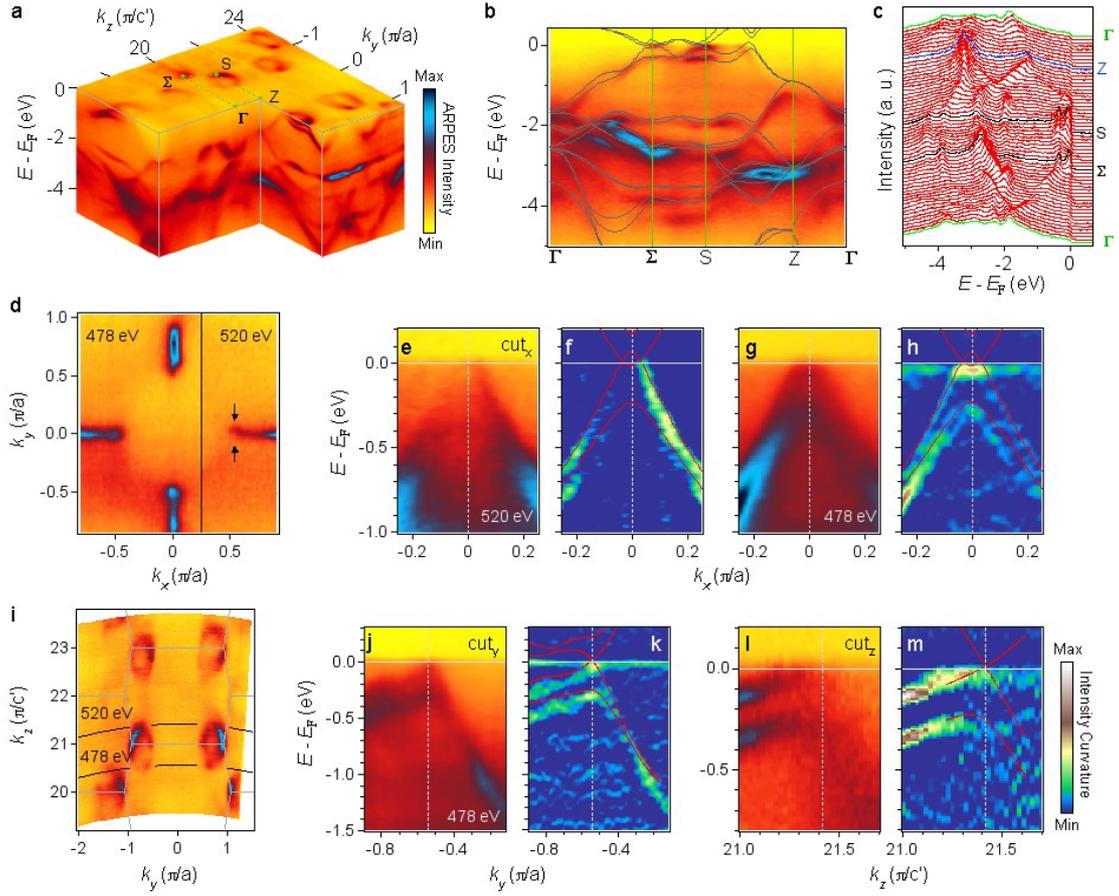

**Figure 2. Bulk electronic structures and W1 type of Weyl cones in TaP. a**, Photoemission intensity of the bulk electronic states as a function of energy relative to Fermi level in the $k_y$-$k_z$ plane. **b-c**, The ARPES spectrum and the energy distribution curves along the high-symmetry lines Γ-Σ-S-Z-Γ. The calculated bands are overlaid on top of the spectrum in **b** for comparison. **d**, Fermi surface map in the $k_x$-$k_y$ plane at $k_z = \pm 0.58\ \pi/c'$, taken with $h\nu$ = 478 and 520 eV. **e-f**, ARPES intensity plot and its curvature plot with calculated bands overlaid on top. The data was acquired with $h\nu$ = 520 eV along $cut_x$ passing through W1. **g-h**, Same as **e-f** but the data was acquired with $h\nu$ = 478 eV. **i**, Fermi surface map in the $k_y$-$k_z$ plane at $k_x = 0$, the data was acquired with soft X-ray. **j-k** and **l-m**, Same as **g-h** but along long $cut_y$ and $cut_z$, respectively.

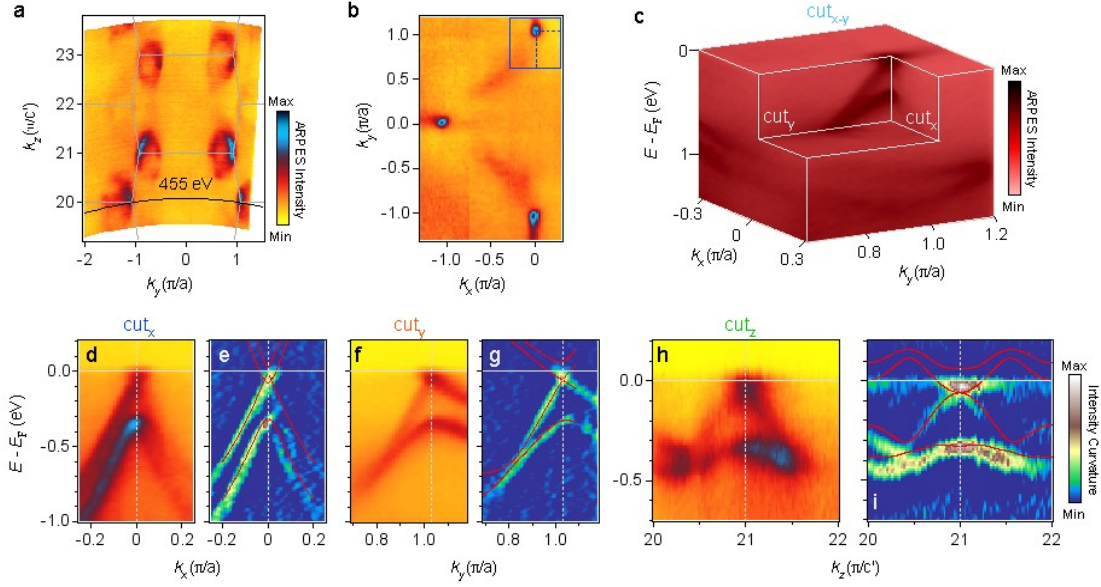

**Figure 3. W2 type of Weyl cones in TaP. a**, Fermi surface map in the $k_y$-$k_z$ plane at $k_x = 0$, acquired with soft X-ray. **b**, Fermi surface map in the $k_x$-$k_y$ plane at $k_z = 0$, taken with $h\nu = 455$ eV as indicated in **a**. **c**, ARPES intensity profile as a function of energy relative to Fermi level in the $k_x$-$k_y$ plane containing a pair of Weyl nodes W2. **d**, Photoemission intensity plot along cut$_x$ passing through W2. **e**, The corresponding curvature plot of the spectrum in **d** with the calculated bands overlaid on top for comparison. **f**-**g** and **h**-**i**, Same as **d**-**e** but along long cut$_y$ and cut$_z$, respectively.

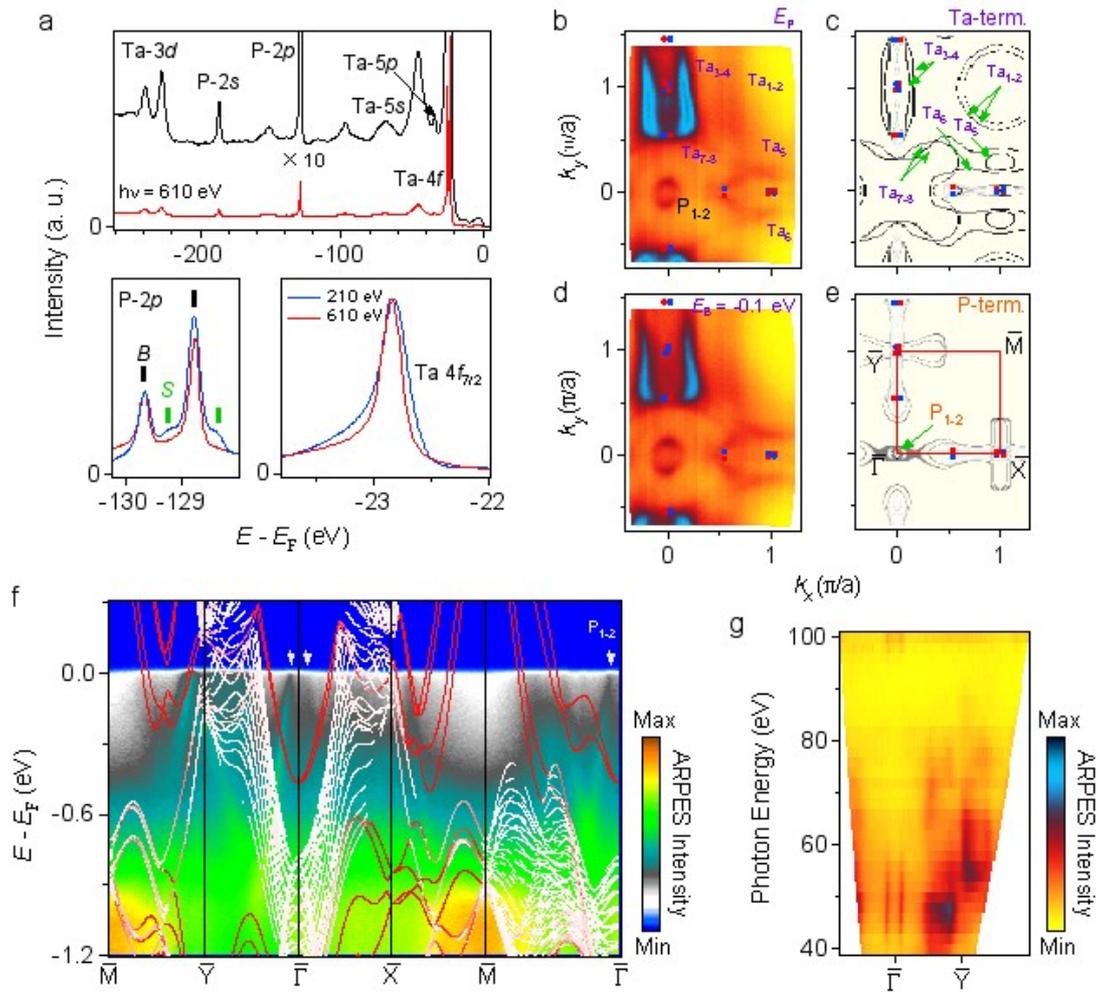

**Figure 4. Fermi arc on the Ta terminated (001) surface of TaP. a**, Core level spectra taken with $h\nu = 610$ eV, and comparisons to the one recorded with VUV light for P-2$p$ and Ta 4$f_{7/2}$ core levels. **b**, Fermi surface map recorded with VUV light ($h\nu = 50$ eV). **c**, The calculated Fermi surface of the surface states on the Ta-terminated (001) surface. **d**, ARPES intensity map at $E_B = -0.1$ eV. **e**, The calculated Fermi surface on P-termination. **f**, ARPES spectra along high symmetry lines acquired with $h\nu = 50$ eV. The calculated bands on Ta termination are overlaid for comparison. The surface bands resulting from the Ta-terminated surface are indicated with color lines. The grey arrows indicate the surface states from P-terminated surface. **g**. Photon-energy-dependent spectrum acquired with $h\nu = 40$-$100$ eV.

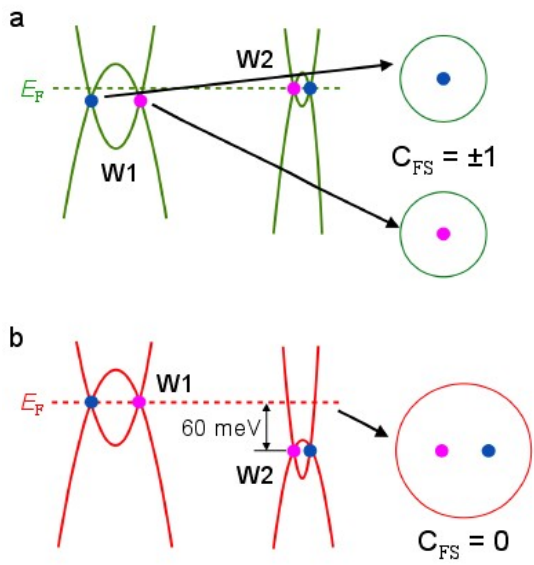

**Figure 5.** **a**, The illustration of energy dispersions along $cut_x$ for W1 and W2 in TaAs. The Fermi surface encloses a single chiral Weyl node and the $C_{FS}$ is ±1. **b**, Same as **a** but for TaP. The Fermi surface encloses two W2 Weyl nodes with opposite chirality and the $C_{FS}$ is zero.